\documentclass[prl,twocolumn,amsmath,amssymb,showpacs,superscriptaddress,floatfix]{revtex4}
\def\vc{\mathbf}
\newcommand{\celc}[0]{^\circ\mathrm{C}}
\newcommand{\kT}[0]{k_\mathrm{B}T}

\usepackage{graphicx}
\usepackage{color}
\usepackage{soul}
\begin{document}
\title{Low-frequency vibrations of soft colloidal glasses}

\author{Ke Chen }
\affiliation{Department of Physics and Astronomy, University of Pennsylvania, Philadelphia, Pennsylvania 19104, USA}

\author{Wouter G. Ellenbroek}
\affiliation{Department of Physics and Astronomy, University of Pennsylvania, Philadelphia, Pennsylvania 19104, USA}

\author{Zexin Zhang}
\affiliation{Department of Physics and Astronomy, University of Pennsylvania, Philadelphia, Pennsylvania 19104, USA}

\author{Daniel T. N. Chen}
\affiliation{Department of Physics and Astronomy, University of Pennsylvania, Philadelphia, Pennsylvania 19104, USA}

\author{Peter Yunker}
\affiliation{Department of Physics and Astronomy, University of Pennsylvania, Philadelphia, Pennsylvania 19104, USA}

\author{Silke Henkes}
\affiliation{Instituut--Lorentz, Universiteit Leiden, Postbus 9506, 2300 RA Leiden, The Netherlands}

\author{Carolina Brito}
\affiliation{Instituut--Lorentz, Universiteit Leiden, Postbus 9506, 2300 RA Leiden, The Netherlands}
\affiliation{Inst. de F\'isica, Universidade Federal do Rio Grande do Sul, CP 15051, 91501-970, Porto Alegre RS, Brazil}

\author{Olivier Dauchot}
\affiliation{Service de Physique de l'\'Etat Condens\'e, CEA-Saclay; URA 2464, CNRS, 91191 Gif-sur-Yvette, France}

\author{Wim van Saarloos}
\affiliation{Instituut--Lorentz, Universiteit Leiden, Postbus 9506, 2300 RA Leiden, The Netherlands}

\author{Andrea J. Liu}
\affiliation{Department of Physics and Astronomy, University of Pennsylvania, Philadelphia, Pennsylvania 19104, USA}

\author{A. G. Yodh}
\affiliation{Department of Physics and Astronomy, University of Pennsylvania, Philadelphia, Pennsylvania 19104, USA}

\date{\today}
\begin{abstract}
We conduct experiments on two-dimensional packings of colloidal thermosensitive hydrogel particles whose packing fraction can be tuned above the jamming transition by varying the temperature.  By measuring displacement correlations between particles, we extract the vibrational properties of a corresponding ``shadow'' system with the same configuration and interactions, but for which the dynamics of the particles are undamped.  The vibrational spectrum and the nature of the modes are very similar to those predicted for zero-temperature idealized sphere models and found in atomic and molecular glasses; there is a boson peak at low frequency that shifts to higher frequency as the system is compressed above the jamming transition.
\end{abstract}
\maketitle

Crystalline solids are all alike in their vibrational properties at low
frequencies; every disordered solid is disordered in its own way.  Disordered solids nonetheless exhibit common low-frequency vibrational properties that are completely unlike those of crystals, which are dominated by sound modes.  
Disordered atomic or molecular solids generically exhibit a ``boson peak," where many more modes appear than
expected for sound.  The excess modes of the boson peak are believed to be responsible for
the unusual behavior of the heat capacity and thermal conductivity at low-to-intermediate temperatures in disordered solids~\cite{pohl02}.  

 It has been proposed that a zero-temperature jamming transition may provide a framework for understanding this unexpected commonality~\cite{liuAR10}.
For frictionless, idealized spheres this jamming transition lies at the threshold of mechanical stability, known as the isostatic point~\cite{liuAR10,epitome}.  As a result of
this coincidence, the vibrational behavior of the marginally jammed solid at
densities just above the jamming transition is fundamentally different from
that of ordinary elastic solids~\cite{silbertPRL05,wyartEPL,xu09,xu09qlarxiv}. 
A new class of low-frequency vibrational modes
arises because the system is at the threshold of mechanical stability~\cite{wyartthesis}; these modes give rise to a divergent boson peak at zero frequency~\cite{silbertPRL05}.
As the system is compressed beyond the jamming transition, the boson peak shrinks in height and shifts upwards in frequency~\cite{silbertPRL05}.
Generalizations of the idealized sphere
model suggest that the boson peaks of a wide class of disordered solids may arise from proximity to the jamming transition~\cite{xu07,wyartthesis,wyart08,wyart08arxiv}.   Moreover, the jamming scenario predicts that systems with larger constituents such as colloids should also have boson peaks.

Colloidal glasses offer signal advantages over atomic or
molecular disordered solids because colloids can be tracked by video microscopy.
Vibrational behavior has been explored in hard-sphere colloids~\cite{ghosh09arxiv} and vibrated granular packings~\cite{brito10arxiv}, but difficulties with statistics~\cite{ghosh09arxiv} or micro-cracks~\cite{brito10arxiv} were encountered.  In contrast, we use deformable, thermosensitive
hydrogel particles to tune the packing fraction {\it in situ}.  Our experiments show unambiguously that the commonality in vibrational properties observed in atomic and molecular glasses extends even to colloidal glasses, in striking confirmation of the jamming scenario.

We study disordered colloidal solids composed of poly($N$-isopropylacrylamide) or NIPA microgel
particles. NIPA particles swell with decreasing temperature, and can thus be tuned at fixed number density from a loose packing
of small particles to a jammed packing of highly deformed, larger particles over the range of a few degrees in temperature.
Particles were loaded between two glass cover slips, creating a monolayer in
which the spheres were confined to move in the horizontal plane.  A binary
mixture (diameters $1\,\mu $m and $1.4\,\mu $m at $T=24.7\celc$, with a large/small number
ratio $\sim0.7$) was employed to suppress crystallization.  The sample was hermetically
sealed using optical glue (Norland 63), and annealed for two hours at
$28\celc$. Data were acquired using standard bright field video microscopy at
temperatures ranging from $24.7\celc$ to $27.2\celc$. Over this
range the thermal energy is essentially constant, so that temperature serves primarily to
change particle diameter and packing fraction. The sample temperature was controlled
by thermal coupling to the microscope objective (BiOptechs), and
the sample was allowed to equilibrate for 15 minutes at each temperature before data acquisition. The region of interest
was far from any boundaries or non-uniform regions within the sample.  The
video rate was 30 frames/s for a duration of 1000 s.  The trajectories of the $N \approx 3600$ particles in the experimental field of
view were extracted using standard particle tracking techniques.  Cage rearrangements did not occur in any of the data sets shown during the 1000 s of run time.  Thus, the system
remained in the same basin of the energy landscape and each particle had a well-defined average position.

\begin{figure}[t]
\includegraphics[width=8.6cm]{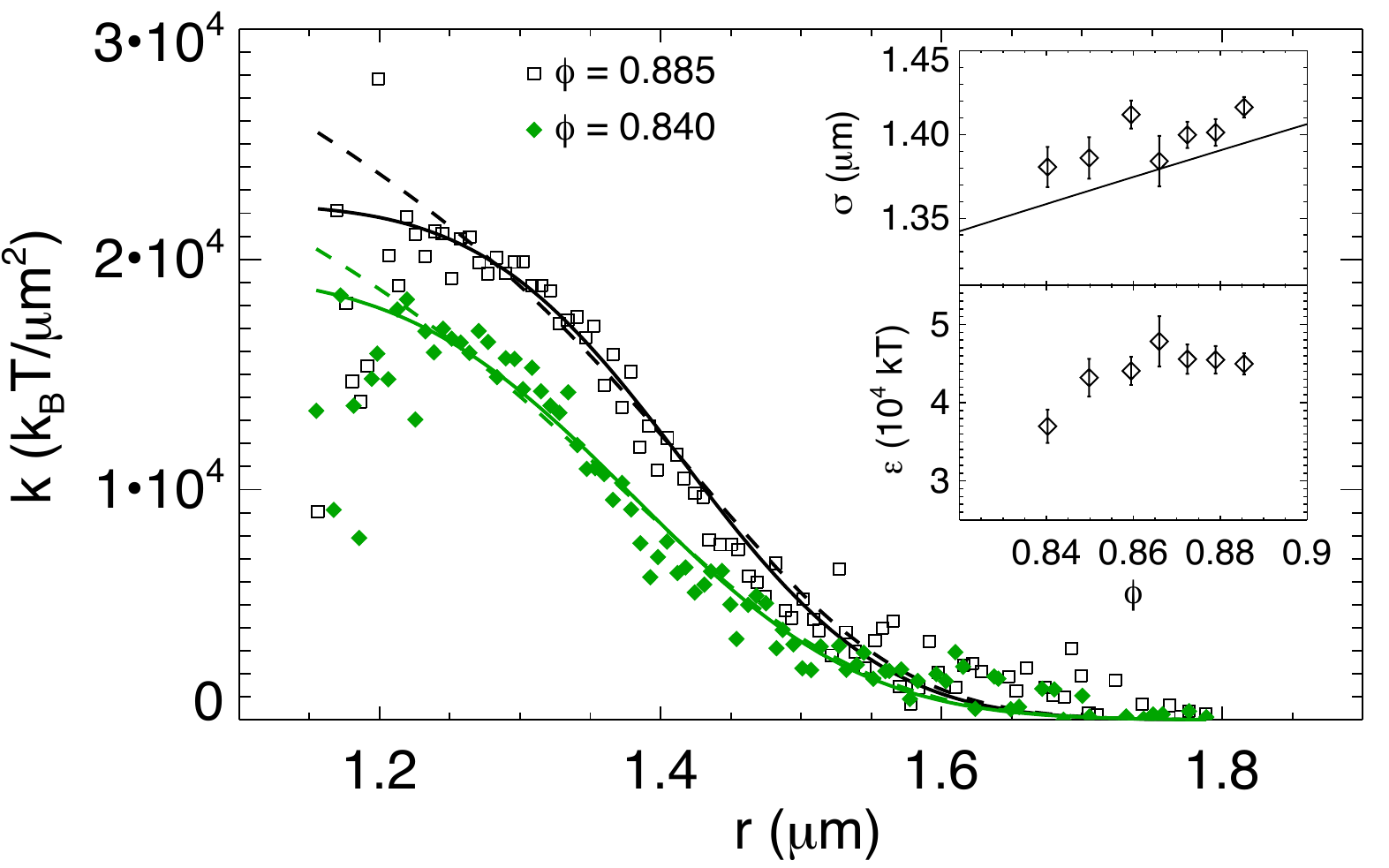}
\caption{(color online) Interaction (in terms of the effective spring constant $k_{ij}$) between
large particles as extracted from the stiffness matrix $K$ at $\phi=0.885$
(squares), and $\phi=0.840$ (diamonds). The results are fit to harmonic (solid
line), and Hertzian (dashed line) interactions, convoluted with a Gaussian
size distribution to take the known polydispersity into account. Inset: Interaction range
$\sigma$ and energy scale $\epsilon$ for the fitted harmonic interaction
as a function of packing fraction. The solid line in the upper inset corresponds to the particle diameter implied by the volume fraction.}
\label{fig:kr}
\end{figure}

As in \cite{ghosh09arxiv,brito10arxiv}, we define $\vc{u}(t)$
as the $2N$-component vector of the displacements of all particles
from their average positions, and extract the time-averaged displacement
correlation matrix, or covariance matrix, as~\cite{ghosh09arxiv,brito10arxiv}
\begin{equation}
\label{Cdef}
C_{ij}=\left\langle u_i(t)u_j(t)\right\rangle_t~,
\end{equation}
where $i,j=1,\ldots,2N$ run over particles and coordinate directions, and the
average runs over time frames.

In the harmonic approximation, the displacement
correlations, $C$, are directly related to the stiffness matrix, $K$, defined as the matrix of
second derivatives of the effective pair interaction potential with respect to particle displacements. To quadratic order, the effective potential energy of the system is
$V=\frac12\vc{u}^\mathrm{T}K\vc{u}$.
Within the energy basin the system is thermally equilibrated, so we can
calculate correlation functions from the partition function, $
Z\propto\int D\vc{u}\exp\left(-{\textstyle\frac12}\beta \vc{u}^\mathrm{T}K\vc{u}\right)$,
where $\beta=1/\kT$.  In particular,
$\langle u_i u_j\rangle=\kT\left(K^{-1}\right)_{ij}$.

\begin{figure}[t]
\includegraphics[width=8cm]{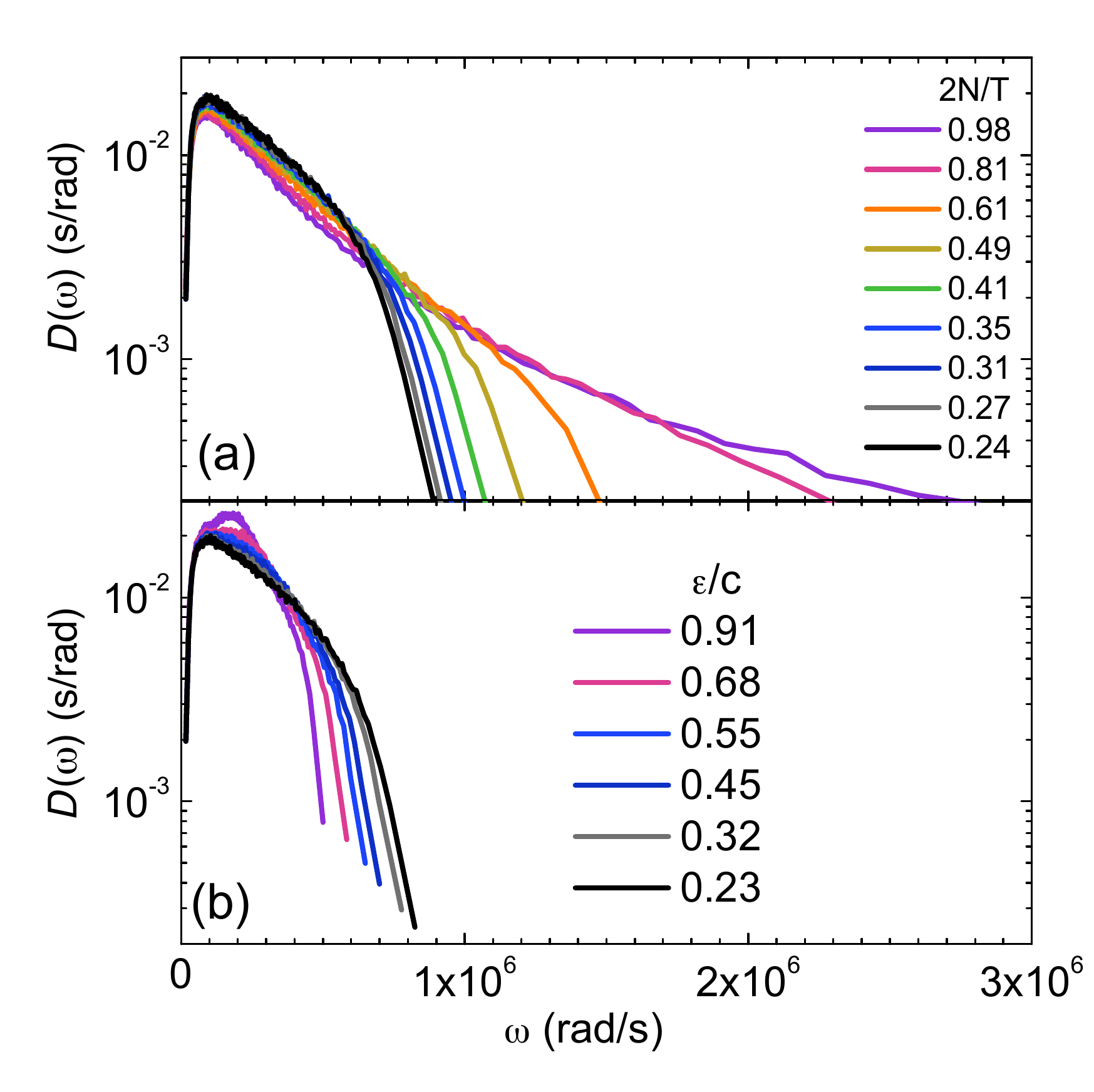}
\caption{(color online)  The vibrational density of states at $\phi=0.859$ averaged over (a) different ratios $2N/T$ of the number of degrees of freedom to the number of frames, as labeled, and (b) different values of the optical resolution relative to the average rms displacement, $\epsilon/c$, as labeled.}
\label{fig:checks}
\end{figure}

We now introduce a \emph{shadow system} of particles thermally equilibrated in the same configuration with the same interactions as the colloids.  In contrast to the colloids, which are suspended in water and whose motion is strongly damped, the virtual particles of the shadow system are undamped.  The real and shadow systems are characterized by the same correlation and stiffness matrices, $C$ and $K$, because these are static equilibrium quantities.  For the shadow system, however, the stiffness matrix is directly related to the dynamical matrix
\begin{equation}
D_{ij}=\frac{K_{ij}}{m_i}=\frac{(C^{-1})_{ij}}{m_i\kT}~.
\end{equation}
whose eigenvectors correspond to the vibrational modes. Thus, this analysis allows direct comparison of the damped colloidal solid to disordered atomic or molecular glasses and to idealized sphere packings.

\begin{figure*}[!t]
\includegraphics[width=\textwidth]{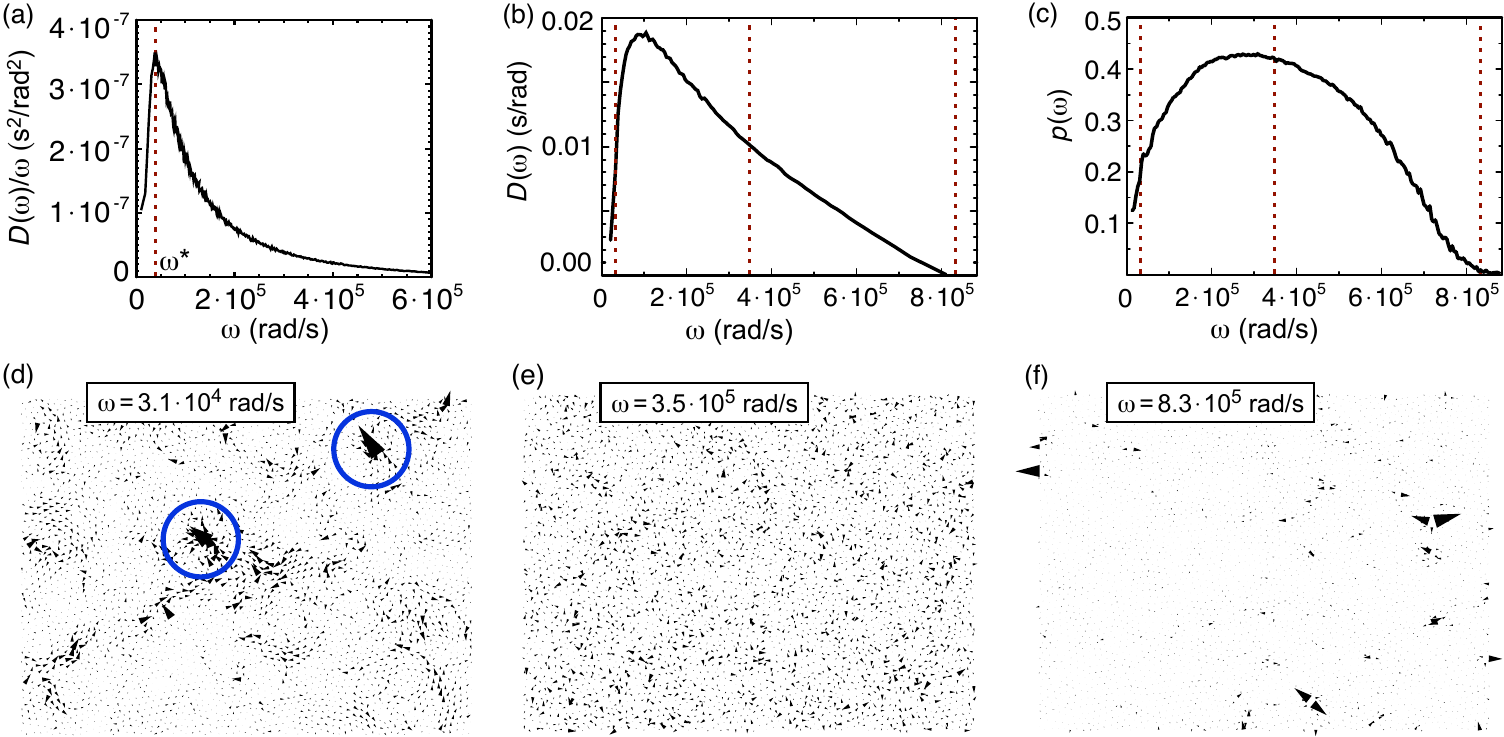}
\caption{(color online) (a) The vibrational density of states relative to the Debye prediction,$D(\omega)/\omega$, for $\phi=0.859$.  The position of the maximum in this plot, shown by the vertical dashed line, defines the boson peak frequency, $\omega^*$.  (b) The density of states, $D(\omega)$  and (c) the participation ratio $p(\omega)$.  The dotted vertical lines in (b) and (c) indicate the frequencies of the modes shown in the remaining panels. (d-f)~Displacement vector plots of
eigenmodes at (d) low, (e) intermediate, and (f) high frequencies. The size
of each arrow is proportional to the displacement of the particle at that position,
weighted by its mass: $|\vc{u}_i|\sqrt{m_i}$.  Blue circles in (d) indicate regions of high-displacements in the quasilocalized mode.}
\label{fig:onephi}
\end{figure*}

One advantage of this analysis is that we can obtain vibrational information about the disordered solid without knowing the effective interaction potential between particles in advance.  As an added bonus, we can extract information about the interactions.
The elements of stiffness matrix $K$ contain direct information
about the second derivative of the effective pair potential,
denoted as $k_{ij}$.  We find that $k_{ij}$ is far above the noise only for adjacent particles.  This observation gives us confidence in the method; clearly, elastic interactions between neighboring particles dominate the data, as they should.  Furthermore, positional information about the large particles is sufficiently accurate to yield the
stiffness as a function of separation, $k(r)$.  We fit $k(r)$ to the second derivative of the harmonic ($\alpha=2$) or
3D Hertzian ($\alpha=5/2$) interaction potentials given by
\begin{equation}
\label{pot}
V(r_{ij})=\frac{\epsilon}{\alpha}\left(1-\frac{r_{ij}}{R_i+R_j}\right)^\alpha\qquad r_{ij}\leq R_i+R_j~,
\end{equation}
and $V=0$ otherwise. We convolute the fitting functions with a Gaussian distribution to account
for the known colloidal particle polydispersity.  The results are plotted in Fig.~\ref{fig:kr}. Note that we cannot differentiate between the harmonic and Hertzian fits; the
difference is tiny except for
extremely large overlaps where statistics are limited. The fit parameters $\epsilon$ and $\sigma=\langle
R_i+R_j\rangle$ for the harmonic potential are shown as insets in
Fig.~\ref{fig:kr}. As one might expect, the average size of the particles
increases with $\phi$, and the energy scale is roughly constant when
$\phi>0.84$.  The solid curve in the top inset of Fig.~\ref{fig:kr} shows the value of the diameter implicit in the assignment of $\phi$ obtained by interpolation of dynamic light scattering data at different temperatures at low concentration.  Here we have measured the elastic contact distance while dynamic light scattering measures a hydrodynamic radius.  The uncertainty in the solid curve is larger than the difference between the curve and our data.  The correlation method therefore provides a useful new way of measuring interparticle interactions at high concentrations.

The accuracy of our results depends on the statistics of the time averages to calculate the matrix elements in Eq.~\ref{Cdef}, as well as on the ability to resolve particle displacements $\vc{u}(t)$.  The optical resolution, $\epsilon \approx 5$nm, at best in our case, must be small compared to the average root-mean-squared displacement, calculated to be $c = \langle C_{ii}^{1/2} \rangle \approx 22$nm at the packing fraction $\phi=0.859$.  Likewise, the number of degrees of freedom, $2N \approx 7200$, should be small compared to the number of time frames, $T=30000$, at most in our case, over which $C_{ij}$ is averaged.  Fig.~\ref{fig:checks}(a) and (b) show the density of vibrational states, $D(\omega)$, as $\epsilon/c$ and $2N/T$ are varied.   $D(\omega)$ is normalized so that $\int_0^\infty D(\omega) d\omega = 2N$, and $\epsilon$ is varied by rounding measured particle displacements.   Note that poor resolution artificially lowers the high end of the spectrum but that inadequate statistics raise the high end, so that the two effects tend to cancel.  The spectrum appears reasonably close to convergence for $T=30000$ and $\epsilon=5$nm, used in the other figures.  The error in the width of the spectrum for these values is estimated as 10\%~\cite{henkesnotes} (see also
\cite{brito10arxiv}). 

The vibrational spectrum for crystals has the Debye form, $D(\omega) \sim \omega$.  In Fig.~\ref{fig:onephi}(a) we plot the measured density of states relative to the Debye prediction, $D(\omega)/\omega$, at $\phi=0.859$, well above the jamming transition at $\phi_c \approx 0.84$~\cite{zhang09} \footnote{We define $\phi_c \approx 0.84$ as the onset of jamming for this system following earlier experiments on the same system~\cite{zhang09}.  At this packing fraction, the peak in the dynamic susceptibility, $\chi_4$, passes out of our time window, and the first peak of the pair correlation function $g(r)$ exhibits a maximum~\cite{zhang09}.}.     For an elastic solid such as a crystal, $D(\omega)/\omega$ would be flat at low frequencies; the presence of a maximum in Fig.~\ref{fig:onephi}(a) indicates the existence of a boson peak and the position of the maximum defines the boson peak frequency, $\omega^*$.   Thus our results show that a boson peak---an important feature in the vibrational spectrum of atomic and molecular glasses---also appears in a disordered colloidal solid. 

Fig.~\ref{fig:onephi}(b) shows the vibrational density of states, $D(\omega)$.  Fig.~\ref{fig:onephi} shows the mode participation ratio, $p(\omega)$, which measures the degree
of spatial localization of a mode $n$: $p(\omega_n) = \left(\sum_i m_i |\mathbf{e}_{n,i}|^2\right)^2/(N\sum_i m_i^2 |\mathbf{e}_{n,i}|^4)$, where ${\bf e}_{n,i}$ is the polarization vector in mode $n$ and $m_i$ is the mass, of particle $i$.  Thus, $p(\omega_n) \sim 1/N$ for a localized mode and $p(\omega_n) \sim {\cal O}(1)$ for an extended mode.  Note that the results are qualitatively the same even if we do not weight the sums with the particle masses.  

At low frequency, Fig.~\ref{fig:onephi}(c) shows that the modes are clearly very different from plane-wave sound modes because they have a low participation ratio; Fig.~\ref{fig:onephi}(d) depicts a typical low-frequency mode.  It is quasi-localized with localized structure superimposed on a plane-wave-like background.  At intermediate frequencies, eigenmodes are highly disordered and extended with $p \sim {\cal O}(1)$~\cite{wyartEPL,silbert09} (Fig.~\ref{fig:onephi}(e)) while at high frequencies the modes are localized, as expected (Fig.~\ref{fig:onephi}(f)).

\begin{figure}[t]
\includegraphics[width=8cm]{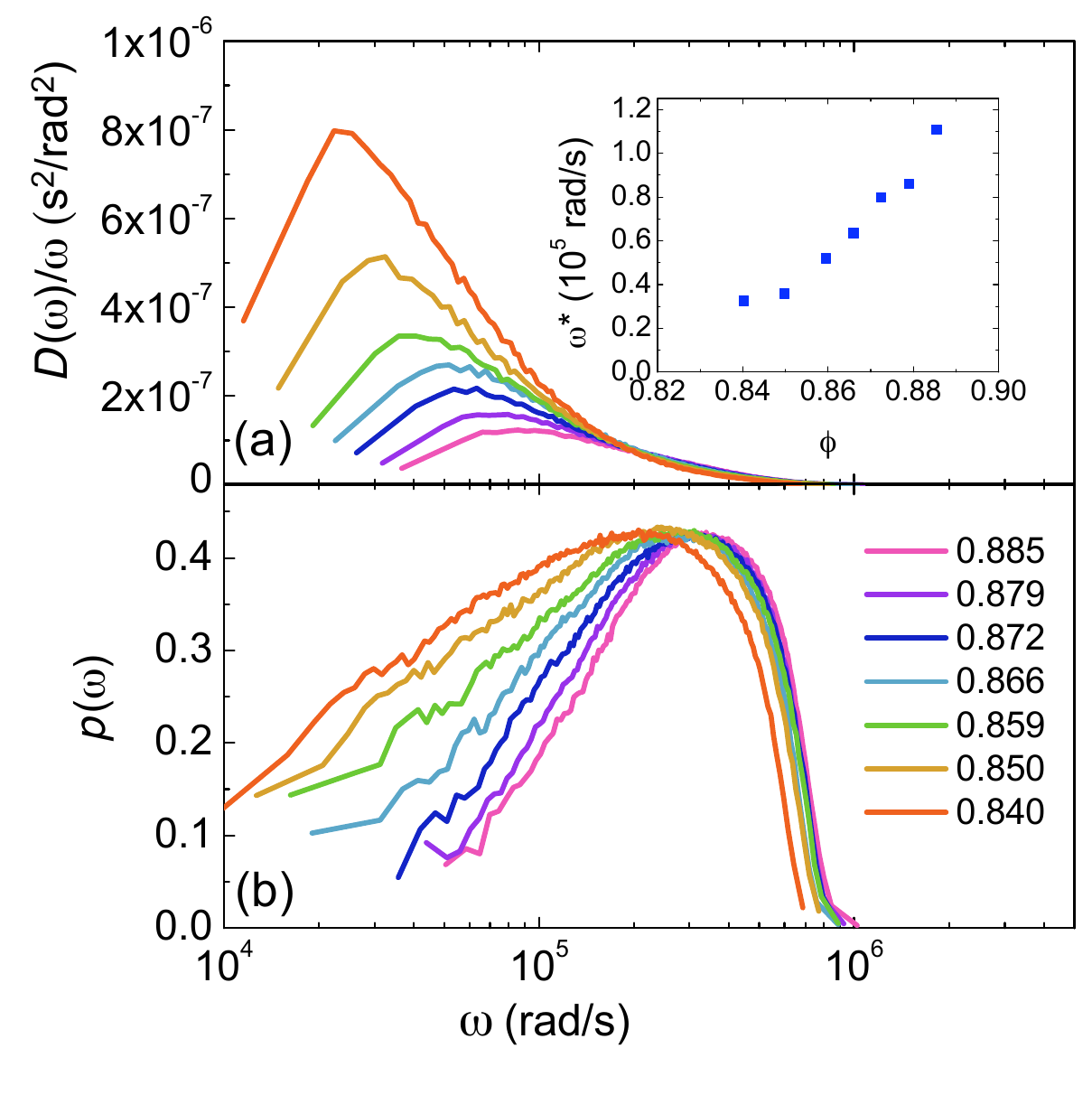}
\caption{ (color online)  (a) $D(\omega)/\omega$ and (b) participation ratio vs.~frequency as a function of packing fraction (as labeled in (b)).   Inset to (a): the
boson peak frequency $\omega^*$ as a function of packing fraction.}
\label{fig:allphi}
\end{figure}

Fig.~\ref{fig:allphi}(a) shows the density of states relative to the Debye prediction, $D(\omega)/\omega$, as a function of  $\phi$ and the boson peak frequency, $\omega^*$ (inset to Fig.~\ref{fig:allphi}(a)).   As expected, $\omega^*$ increases with $\phi$~\cite{silbertPRL05}.  It does not vanish at $\phi_c$ because the system is at nonzero temperature.  Similar thermal smearing of the zero-temperature jamming transition is observed in idealized spheres~\cite{xu09tharxiv}.

Fig.~\ref{fig:allphi}(b) shows that the shape of $p(\omega)$ is similar at all packing fractions.  As the system is decompressed towards $\phi_c$, however, the low-$\omega$ quasilocalized modes shift down in frequency, edging closer to instability~\cite{xu09qlarxiv}.

Quasi-localized modes at low frequency have been observed in idealized
sphere packings as well as in atomic or molecular glasses (see ~\cite{xu09qlarxiv} and references therein).  In the sphere packings, these modes are associated with the lowest energy barriers for
rearrangements~\cite{xu09qlarxiv} and are the ones that tend to go
unstable.  It has been shown that under mechanical load, a mode shifts downwards in frequency to zero, signaling a rearrangement~\cite{maloney06}.  Such modes are quasi-localized, and the rearrangement occurs in the region in which the mode has high displacements (circled in Fig.~\ref{fig:onephi}(d))~\cite{lisaprivate}.   Low-frequency quasi-localized modes are also
connected to irreversible particle rearrangements driven by thermal fluctuations above the glass transition
temperature~\cite{widmercooper08,widmercooper09}.    Thus, our experiments demonstrate that it is possible, in a real disordered solid for which we do not know the interparticle interactions {\it a priori}, to identify localized regions that are likely to be predisposed towards failure.

We thank N. Xu, D. Bonn, J. Crocker, D. J. Durian and S. R. Nagel for helpful discussions.  This work was funded by DMR 080488 (AGY), PENN-MRSEC DMR-0520020 (KC and WE), NASA NNX08AO0G (AGY) and DOE DE-FG02-05ER46199 (AJL and WE).


\begin{thebibliography}{10}
\providecommand{\url}[1]{\texttt{#1}}
\providecommand{\urlprefix}{URL }
\providecommand{\eprint}[2][]{\url{#2}}

\bibitem{ashcroftmermin}
N.~W. Ashcroft and D.~N. Mermin, \emph{Solid State Physics} (Thomson Learning,
  Toronto, 1976).

\bibitem{pohl02}
R.~O. Pohl, X.~Liu, and E.~Thompson, Rev. Mod. Phys. \textbf{74}, 991 (2002).

\bibitem{liuAR10}
A.~J. Liu and S.~R. Nagel, Ann. Rev. of Cond. Mat. Phys. \textbf{1}, in press
  (2010).

\bibitem{epitome}
C.~S. O'Hern \emph{et~al.}, Phys.\ Rev.\ E \textbf{68}, 011306 (2003).

\bibitem{silbertPRL05}
L.~E. Silbert, A.~J. Liu, and S.~R. Nagel, Phys.\ Rev.\ Lett. \textbf{95},
  098301 (2005).

\bibitem{wyartEPL}
M.~Wyart, S.~R. Nagel, and T.~A. Witten, Europhys.\ Lett. \textbf{72}, 486
  (2005).

\bibitem{xu09}
N.~Xu \emph{et~al.}, Phys. Rev. Lett. \textbf{102}, 038001 (2009).

\bibitem{xu09qlarxiv}
N.~Xu \emph{et~al.}, arXiv:0909.3701v1 (2009).

\bibitem{xu07}
N.~Xu \emph{et~al.}, Phys. Rev. Lett. \textbf{98}, 175502 (2007).

\bibitem{wyartthesis}
M.~Wyart, Ann.\ Phys.\ Fr. \textbf{30, No.\ 3}, 1 (2005).

\bibitem{wyart08}
M.~Wyart \emph{et~al.}, Phys.\ Rev.\ Lett. \textbf{101}, 215501 (2008).

\bibitem{wyart08arxiv}
M.~Wyart, arXiv:0806.4596 (2008).

\bibitem{ghosh09arxiv}
A.~Ghosh \emph{et~al.}, arXiv:0910.3231v1 (2009).

\bibitem{brito10arxiv}
C.~Brito \emph{et~al.}, arXiv:1003.1529v1 (2010).

\bibitem{henkesnotes}
S.~Henkes \emph{et~al.}, unpublished (2010).

\bibitem{zhang09}
Z.~Zhang \emph{et~al.}, Nature \textbf{459}, 230 (2009).

\bibitem{silbert09}
L.~E. Silbert, A.~J. Liu, and S.~R. Nagel, Phys.\ Rev.\ E \textbf{79}, 021308
  (2009).

\bibitem{xu09tharxiv}
N.~Xu, arXiv:0911.1576 (2009).

\bibitem{maloney06}
C.~E. Maloney and A.~Lema\^\i{}tre, Phys. Rev. E \textbf{74}, 016118 (2006).

\bibitem{lisaprivate}
M.~E. Manning and A.~J. Liu, private comm. (2009).

\bibitem{widmercooper08}
A.~Widmer-Cooper \emph{et~al.}, Nat.\ Phys. \textbf{4}, 711 (2008).

\bibitem{widmercooper09}
A.~Widmer-Cooper \emph{et~al.}, J. Chem.\ Phys. \textbf{131}, 194508 (2009).

\end{thebibliography}
\end{document}